\documentclass[conference]{IEEEtran}
\pdfoutput=1
\IEEEoverridecommandlockouts
\usepackage{cite}
\usepackage{amsmath,amssymb,amsfonts}
\usepackage{graphicx}
\usepackage{textcomp}
\usepackage{xcolor}
\usepackage{subfig}
\usepackage[T1]{fontenc}

\begin{document}

\title{Single Plane-Wave Imaging using Physics-Based Deep Learning\\
\thanks{This work was supported by CWI, the Dutch Research Council (NWO 613.009.106) and Applus+ RTD. } }

\author{
    \IEEEauthorblockN{Georgios Pilikos\IEEEauthorrefmark{1}, Chris L. de Korte\IEEEauthorrefmark{2}, Tristan van Leeuwen\IEEEauthorrefmark{1} and Felix Lucka\IEEEauthorrefmark{1}}
    \IEEEauthorblockA{\IEEEauthorrefmark{1}Computational Imaging, Centrum Wiskunde \& Informatica, Amsterdam, NL
    }
    \IEEEauthorblockA{\IEEEauthorrefmark{2} Department of Radiology and Nuclear Medicine, Radboud University Medical Center, Nijmegen, NL  }
}

\maketitle

\begin{abstract}
In plane-wave imaging, multiple unfocused ultrasound waves are transmitted into a medium of interest from different angles and an image is formed from the recorded reflections. The number of plane waves used leads to a trade-off between frame-rate and image quality, with single-plane-wave (SPW) imaging being the fastest possible modality with the worst image quality. Recently, deep learning methods have been proposed to improve ultrasound imaging. One approach is to use image-to-image networks that work on the formed image and another is to directly learn a mapping from data to an image. Both approaches utilize purely data-driven models and require deep, expressive network architectures, combined with large numbers of training samples to obtain good results. Here, we propose a data-to-image architecture that incorporates a wave-physics-based image formation algorithm in-between deep convolutional neural networks. To achieve this, we implement the Fourier (FK) migration method as network layers and train the whole network end-to-end. We compare our proposed data-to-image network with an image-to-image network in simulated data experiments, mimicking a medical ultrasound application. Experiments show that it is possible to obtain high-quality SPW images, almost similar to an image formed using 75 plane waves over an angular range of $\pm$16$^\circ$. This illustrates the great potential of combining deep neural networks with physics-based image formation algorithms for SPW imaging.
\end{abstract}


\section{Introduction}
Ultrasound imaging involves the transmission of waves into a medium and the reception of the resulting acoustic wavefield. Efficient data acquisition is essential in order to enable very fast imaging. This is necessary for certain medical applications, such as shear-wave elastography and blood flow mapping \cite{ultrafast}. Conventional medical ultrasound uses focused beams to form images. In recent years, unfocused, plane-wave transmission is being utilized to obtain higher frame rates \cite{stolts} \cite{chris}. Nevertheless, obtaining high-quality images from unfocused waves is challenging. This can be improved by using many plane-wave transmissions, with a wider coverage of angles. By coherently compounding multiple steered plane waves, it is possible to compute high-quality images, however, with more transmissions, data acquisition becomes slower. 

There have been many efforts to form high-quality images from a few plane waves, e.g., the Delay-And-Sum (DAS) image formation algorithm was adapted to plane-wave imaging and combined with coherent compounding \cite{montaldo}. The computational speed with which image formation algorithms operate can present another challenge. Fast run-time is essential to not slow down the ultrasound processing pipeline. This sparked an interest in Fourier domain methods that apply the fast Fourier transform (FFT) for faster execution \cite{fft1}. Fourier-based imaging relies on a spectral re-mapping from the Fourier spectrum of the data to the spectrum of the image. In particular, the Stolt's FK migration, referred to hereafter as Fourier (FK) migration, adapted for ultrasound plane-wave imaging \cite{stolts}, has been very popular. However, when using only a single plane-wave transmission, image quality is poor. 

Recently, deep learning has entered the ultrasound imaging field, achieving great results in many applications \cite{dlus}. Neural networks are being used to estimate images \cite{byram1}, using ultrasound data as input and trained in order to beamform an image as their prediction \cite{acb} \cite{perdios}. Further work has been performed to obtain an image and a corresponding material segmentation simultaneously from single channel raw data \cite{segmNair}. These data-to-image networks aim at learning an approximate inverse of the image-to-data mapping. Without an explicit image formation step in the network architecture that links ultrasound data points with image locations, these networks have to be expressive (deep) and their training needs large numbers of training samples. Alternative deep learning techniques have been proposed that combine the great expressive power of data-driven neural networks with traditional image formation techniques. In particular, deep learning and the DAS algorithm have been combined in an end-to-end deep learning framework, obtaining superior results compared to purely data-driven models for both imaging and segmentation \cite{pil1} \cite{pil2}.

\begin{figure*}
\centering
\includegraphics[scale=0.53]{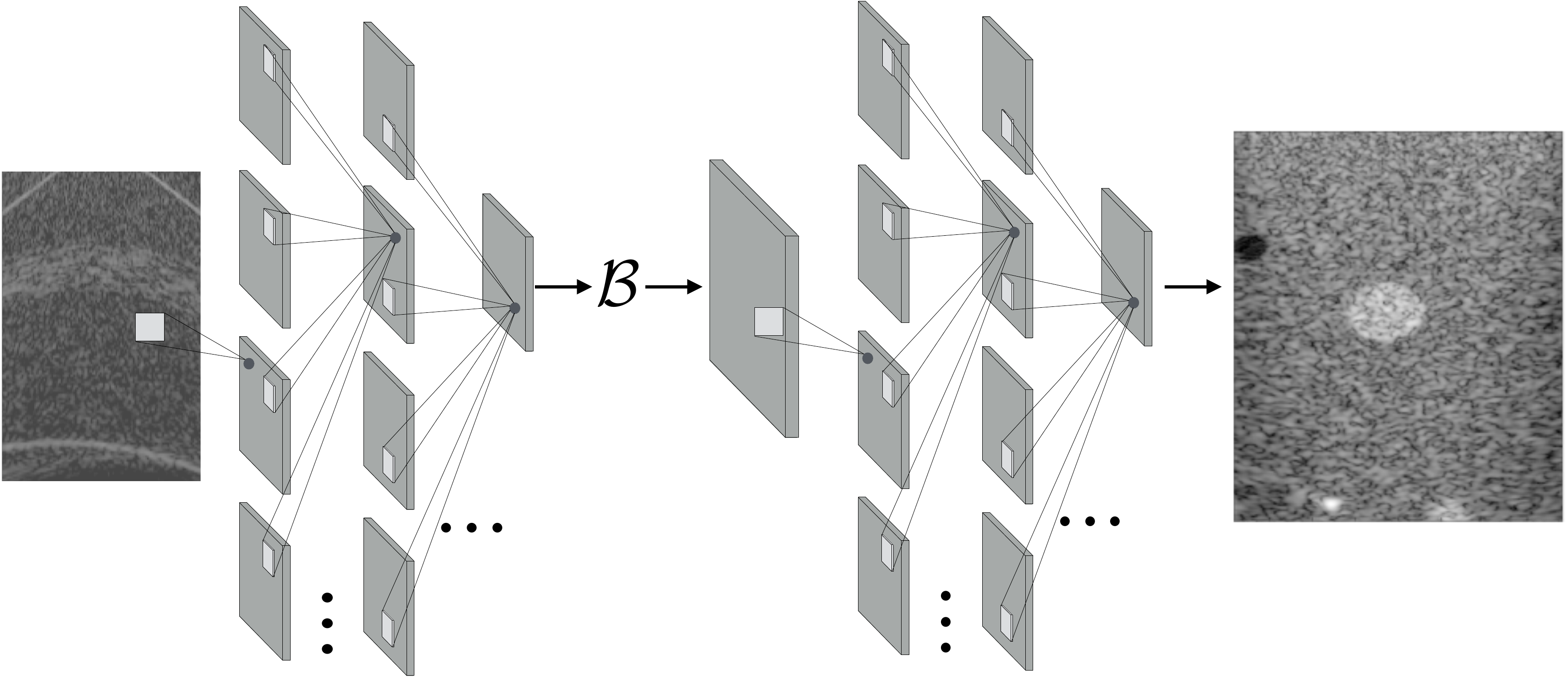}
\caption{A 2D DCNN receives single plane-wave data and performs optimal data pre-processing prior to image formation. FK migration, $\mathcal{B}$, is incorporated within training to compute intermediate images using a traditional physics-based image formation algorithm. Then, another 2D DCNN is used to perform image post-processing to obtain a final, enhanced image. One filter at one location per layer is shown, with the depth and width of layers included only for illustration purposes.}
\label{1}
\end{figure*}

Here, we propose a novel deep learning architecture for single plane-wave imaging. Single plane-wave data are processed by our method and an improved final image is estimated. We implement the FK migration algorithm as network layers and incorporate it in-between two deep convolutional neural networks as illustrated in Figure \ref{1}. We demonstrate that by including the FK migration into data-to-image networks, we obtain high-quality images from a single plane wave with only a few hundred training samples. In section 2, we describe the plane-wave imaging setup and the FK migration algorithm. In section 3, we provide further details about our proposed end-to-end deep learning method. In section 4, we perform experiments on synthetic plane-wave data mimicking a medical ultrasound application, and compare with an image-to-image network. Finally, in section 5, we discuss and conclude our work.

\begin{figure}
\centering
\includegraphics[scale=0.19]{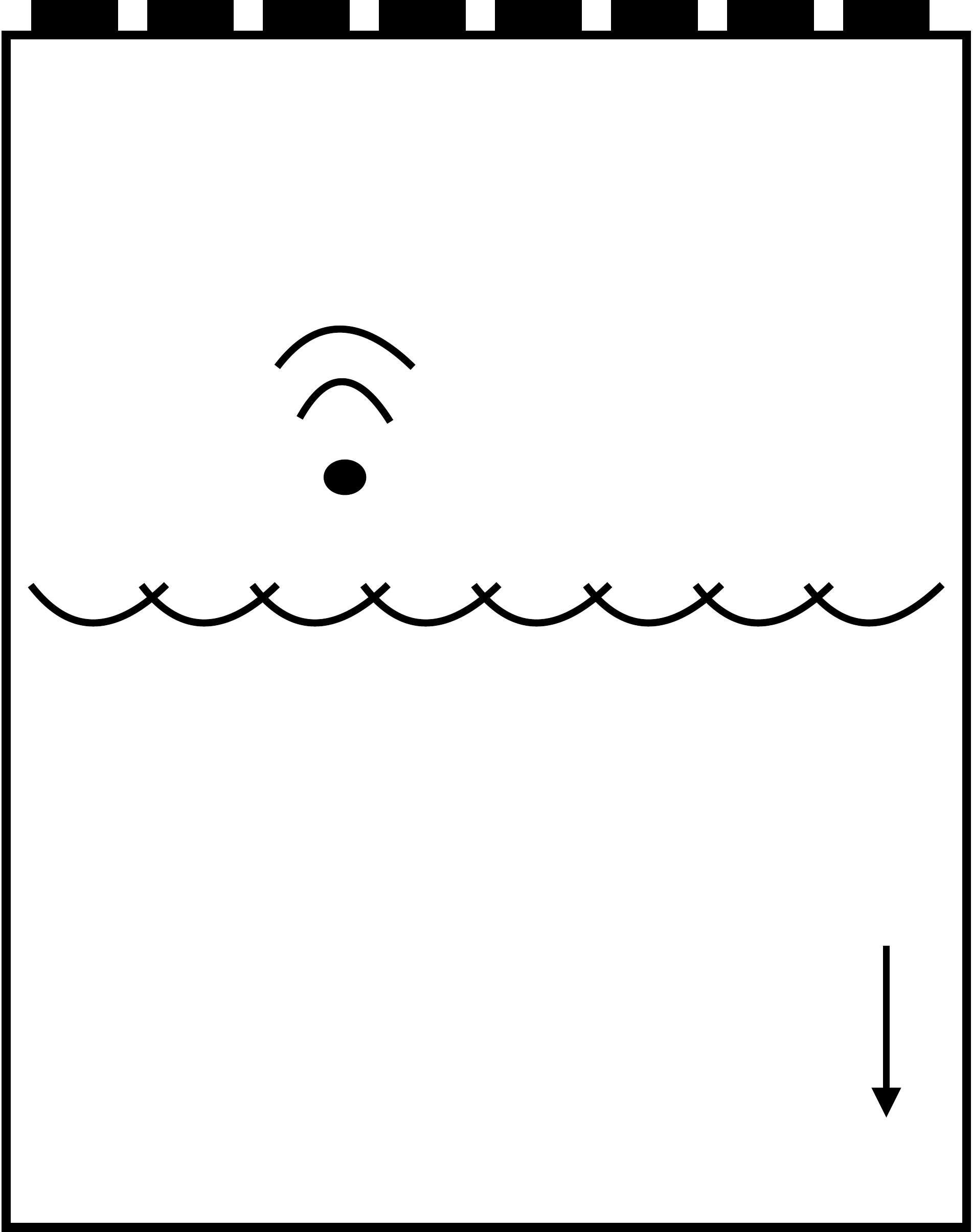}
\caption{Plane-wave transmission by firing all elements of a linear array at the boundary of the domain. Inclusions act as sources when plane-waves pass through them, reflecting data back to the linear array. Delaying the firing of certain elements can steer the plane-wave, enabling plane-wave transmission of different angles, capturing data from different viewpoints and improving the final image formation.}
\label{pw}
\end{figure}

\section{Plane-Wave Imaging and the Fourier Migration}
Plane-wave transmission is achieved by firing all elements of a linear array as illustrated in Figure \ref{pw}. By delaying the firing of elements in a pre-determined manner, it is possible to steer the plane wave in a certain transmission angle. This enables the insonification from different angles and provides complementary information that can improve the image formation. Reflectors within the domain act as sources and waves travel back to the linear array which receives the data. 

The received data are processed by an image formation method to create images. Here, we examine the Fourier (FK) migration which is a widely deployed technique in the ultrasound plane-wave imaging community. It uses the Stolt's spectral re-mapping \cite{stolts} and the physical assumption of the exploding reflector model, where reflector sources within the medium are assumed to be exploded simultaneously. Using the wave equation, the location and intensity of the reflector sources could be reconstructed from data received at the boundaries of the domain.

First, the data are Fourier transformed in the axial dimension. This is then phase-shifted to account for signal delays (such as delays for different transmission angles) and then another Fourier transform is performed in the lateral dimension. The resulting Fourier domain of the plane-wave data is mapped to the Fourier domain of the image using spectral re-mapping. This is implemented using linear interpolation. Finally, an inverse Fourier transform produces the migrated image. Further information on the FK migration, its relation to the Helmholtz equation and implementation details can be found in \cite{stolts} and \cite{chris}. 
\begin{figure*}
\centering
\includegraphics[scale=0.21]{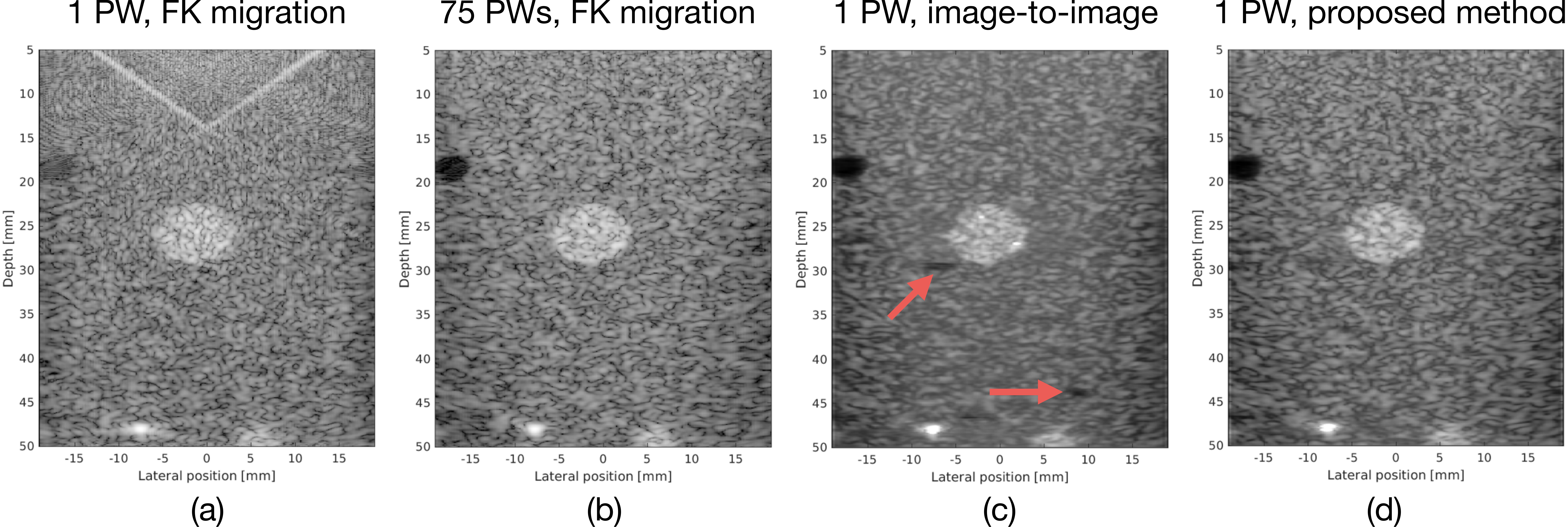}
\caption{(a) Image from 1 plane wave (PW) using FK migration (PSNR = 19.39 dB), (b) image from 75 plane waves using FK migration, (c) image from 1 plane wave with an image-to-image network (PSNR = 21.56 dB), (d) image from 1 plane wave using our proposed method (PSNR = 23.87 dB). The red arrows in (c) illustrate two anechoic cysts that are predicted incorrectly by the image-to-image network. On the other hand, (d) illustrates the prediction of our proposed method, producing a high-quality image.}
\label{2}
\end{figure*}
\section{Proposed Deep FK Migration}
In this work, we implement all steps of the FK migration algorithm as differentiable network layers. In principle, all steps correspond to linear mappings, so all it takes to turn them into differentiable programs that allow for error back propagation is to derive and implement the corresponding adjoint mappings. The entire FK migration is summarized as,
\begin{equation}
\mathbf u = \mathcal{B} \mathbf f,
\end{equation} where $\mathbf u$ is the computed image, $\mathcal{B}$ is a linear operator, referred to as FK migration operator hereafter and $\mathbf f$ is the plane-wave data. The FK migration operator contains an approximation of the underlying wave physics and is utilized in our proposed physics-based deep learning architecture.

In order to obtain images from plane-wave data, we need to estimate an approximate inverse of the image-to-data mapping. Image formation algorithms and deep-learning approaches are being used for this. The first category utilizes approximations of traditional physics-based models and the second employs large amounts of training data to learn approximate models of the underlying wave physics. In this work, we propose an architecture that exploits both categories of models to obtain high-quality images from single plane-wave data using only a relatively small number of training samples.

First, a 2D Deep Convolutional Neural Network (DCNN) is used for data pre-processing. This is a data-to-data mapping which receives single plane-wave data and learns convolutional filters for optimal data processing prior to image formation. This is followed by the FK migration operator, $\mathcal{B}$. It receives the pre-processed data and creates an intermediate image. Then, another 2D DCNN is implemented as an image post-processing step. This is an image-to-image mapping that learns optimal convolutional filters to produce an enhanced final image. Each DCNN has 8 layers with 64 channels per layer.  Weight standardization \cite{ws} and group normalization \cite{gn} is used per layer for training stability since we are only using one training sample per mini-batch. Figure \ref{1} illustrates our proposed physics-based deep learning architecture.

\section{Experiments}
In order to evaluate our proposed deep learning architecture, we use single plane-wave imaging with simulated data mimicking a medical ultrasound application. The wave simulations use speed-of-sound maps that model speckle and various types of inclusions within the medium of interest. We simulate 75 angles of plane-wave data, with angles covering the angular range of $\pm 16^\circ$ with uniform spacing. Speckle, hyper-echoic, anechoic cysts and Nylon fibre inclusions are numerically simulated. The inclusions were varied randomly in location, size and amount to create different scenarios. The background speed of sound was set to $1540$ m/s. The anechoic regions were set as the background and the hyper-echoic regions were randomly varied to have speed-of-sound different to the background speed-of-sound. The Nylon fibre inclusions were set to have speed-of-sound equal to $2620$ m/s. An example can be seen in Figure \ref{2}(b) where an optimal image is composed from 75 angles, coherently compounded using FK migration.

We simulated the plane-wave data using the k-Wave toolbox \cite{kwave} and used the Fourier migration algorithm to produce images. In total, we generated 300 random scenarios, from which 280 were used as training data and 20 as test data. For our proposed architecture, plane-wave data from one angle and a Fourier migrated image from 75 angles are used as pairs for training. In order to evaluate our proposed deep learning method, we also examined another architecture used in the field. We trained an image-to-image network which uses only one 2D DCNN, which is similar to our proposed architecture but without the data-to-data network in the beginning. However, the image-to-image network that we used for comparison, has the same total expressive power as our proposed architecture, i.e. 16 layers with 64 channels per layer. All images are assumed to be Fourier migrated using plane-wave data from one angle. Thus, they need to be improved to an image quality comparable to an image that has been Fourier migrated using plane-wave data from 75 angles. Both architectures are implemented in PyTorch \cite{pyT} using the Adam optimization \cite{adam} with $0.01$ as a learning rate, to optimize network parameters.

Figure \ref{2}(a) includes an image computed from a single, zero-angled, plane wave using FK migration. This is used as input to the image-to-image network that we compare with. Figure \ref{2}(b) shows an image computed from 75 plane waves of various steering angles using FK migration which was used as ground truth. Figure \ref{2}(c) shows the image computed from a single, zero-angled plane wave using an image-to-image network. We can see that the image-to-image network wrongly predicts two small anechoic regions and blurs the image. Figure \ref{2}(d) shows a high-quality image computed from a single, zero-angled plane wave using our proposed method. However, it blurs speckle and reduces image contrast. Overall, visually, our proposed method obtains an accurate image, which is comparable to the ground truth of this scenario.

For a quantitative comparison, the Peak-Signal-to-Noise-Ratio
(PSNR) was estimated on 20 test samples. The average PSNR of our proposed approach using a single plane wave is 23.17 dB as opposed to 21.57 dB for the image-to-image network. The average PSNR of the FK migration algorithm without deep learning is 18.07 dB. This illustrates the potential improvements that can be achieved for single plane-wave imaging and the superiority of our proposed data-to-image network, with the FK migration included within deep learning.

\section{Discussion and Conclusion}
Plane-wave imaging enables fast ultrasound data acquisition that is necessary for certain applications. However, using traditional image formation methods, such as the Fourier migration and Delay-And-Sum algorithms, multiple plane-wave transmissions are required in order to obtain high-quality images. Deep learning has been proposed to improve ultrasound images. In this work, we proposed to combine both deep-learning methods and traditional physics-based image formation techniques. We introduced a novel architecture where the Fourier migration steps are implemented as network layers in-between deep convolutional neural networks. These learn to optimally pre-process plane-wave data and post-process intermediate images to produce a final, improved image. We trained our proposed method using simulated data and used a single plane wave as input. The prediction of our method is an improved image that is comparable to one that has been Fourier migrated using 75 angles of plane-wave data. Experiments have shown that our proposed method, which receives plane-wave data, outperforms an image-to-image network that only processes images. Overall, our method is able to locate inclusions accurately, however, it blurs speckle and the image contrast is reduced. Further experiments and tests are required to determine the limitations of our method using more realistic scenarios. In general, it has been shown that combining deep learning and traditional, physics-based image formation can produce high-quality images using limited training data. This illustrates the great potential of our proposed method for single plane-wave imaging.

\section{Acknowledgements}
We would like to thank Gijs Hendriks from Radboud University Medical Center for useful discussion on ultrasound plane-wave imaging.

\bibliographystyle{IEEEtran}

\bibliography{references}

\end{document}